\documentclass[runningheads]{llncs}
\usepackage{graphicx}
\usepackage{multirow}
\usepackage{{booktabs}}
\usepackage{xurl}
\usepackage{amsmath}
\usepackage{amssymb}
\DeclareMathOperator{\E}{\mathbb{E}}
\usepackage{float} 

\begin{document}

\title{[RE] Modeling Personalized Item Frequency Information for Next-basket Recommendation}
\titlerunning{ [RE] Modeling PIF Information for NBR}

\author{Sławomir Garcarz\inst{1, 2} \and
Avik Pal\inst{1, 2} \and
Pim Praat\inst{1, 2}}

\institute{University of Amsterdam  \and
All authors contributed equally}

\maketitle              

\begin{abstract}
This paper focuses on reproducing and extending the results of the paper: \textit{``Modeling Personalized Item Frequency Information for Next-basket Recommendation"} which introduced the TIFU-KNN model and proposed to utilize \textit{Personalized Item Frequency} (PIF) for Next Basket Recommendation (NBR). We utilized publicly available grocery shopping datasets used in the original paper and incorporated additional datasets to assess the generalizability of the findings. We evaluated the performance of the models using metrics such as Recall@K, NDCG@K, personalized-hit ratio (PHR), and Mean Reciprocal Rank (MRR). Furthermore, we conducted a thorough examination of fairness by considering user characteristics such as average basket size, item popularity, and novelty. Lastly, we introduced novel $\beta$-VAE architecture to model NBR. The experimental results confirmed that the reproduced model, TIFU-KNN, outperforms the baseline model, Personal Top Frequency, on various datasets and metrics. The findings also highlight the challenges posed by smaller basket sizes in some datasets and suggest avenues for future research to improve NBR performance.

\end{abstract}

\section{Introduction}
Applications of \emph{recommender systems} \cite{RecsysCharu16} have been widespread recently as organizations look to promote and sell their products using state-of-the-art tools and algorithms. Next-basket recommendation (NBR) is a problem within this field that aims to predict a subsequent set of items for a user and is prevalent in the e-commerce and retail industries. Hu et al. \cite{modeling-pif-hu} showed that the user's item purchase frequency information modeled as \emph{Personalized Item Frequency} (PIF) vectors provide two useful signals for the NBR task - \emph{repeated} and \emph{collaborative} purchase patterns. They propose a method to capture the two patterns along with the temporal dynamics of previous baskets using a simple k-nearest neighbors (kNN) approach called temporal-item-frequency-based user-KNN (TIFU-KNN). They demonstrate its effectiveness on four \emph{real-world} datasets using recall and NDCG evaluation metrics.

In this work, we verify the reproducibility of the proposed TIFU-KNN model by reproducing the results on the four datasets. We also keep the Personalized Top Frequency method \cite{modeling-pif-hu} as our baseline because it's rarely beaten as repeated purchase patterns dominate in most real-world instances. It's also observed that the datasets used are all related to grocery shopping and might not provide a broad view of the method covering the task of NBR. Hence, we test the TIFU-KNN method on additional NBR datasets to examine its robustness. To further characterize the method in comparison to the baseline frequency method, we also evaluate the method on other NBR metrics - MRR, and PHR. Additionally, we observe a lack but a definite need for fairness metrics in the literature for NBR. Hence, we also introduce metrics to validate and visualize the amount of fairness the NBR methods exhibit towards both users and items. 

Finally, we also try to model the sparse user representations in a latent space with a deep model using a Variational Autoencoder (VAE) architecture \cite{KingmaW13} to capture intrinsic patterns of the user PIF vectors. We also experiment with kNN approach on these dense vectors produced by VAE to inspect the importance of collaborative purchase patterns for NBR.

Thus, we ask the following research questions in our study:
\begin{itemize}
    \item RQ1: Are the results for TIFU-KNN from the original paper \cite{modeling-pif-hu} reproducible?
    \item RQ2: Is TIFU-KNN robust enough to perform well on other NBR datasets? 
    \item RQ3: Is TIFU-KNN performance consistent on metrics beyond recall and NDCG?
    \item RQ4: Is the TIFU-KNN method fair towards users and items?
    \item RQ5: Can a deep model like VAE better capture the purchase patterns?
\end{itemize}

In summary, our experiments show that the results of TIFU-KNN in the original paper are reproducible with little variations. The model again outperforms plain frequency methods on new datasets which is also verified using additional metrics. Our fairness analysis proves that model performance is dependent on the dataset and user characteristics. Lastly, NBR with VAE architecture remains challenging but still provides some useful trends, and future work in better representation learning could further provide gainful insights into recommender systems.

\section{Methodology} 

In this section, we will begin by defining both the models presented in the original paper and our extension, which incorporates a VAE model. We will then discuss the datasets used, the metrics employed, and our definitions of fairness in different aspects for the experiments.

\subsection{Models}
\subsubsection{Personal Top Frequency}
uses the PIF information directly, as it predicts the most frequent items appearing in past baskets for a user directly as the predicted basket. It is considered the dominant baseline in the field, due to its effectiveness on most datasets and metrics. A downside, however, as it only considers historical records from the user, is that it does not present unseen items to the target user \cite{li2023basket}. We also refer to this method as \textbf{TopPersonal}.

\subsubsection{TIFU-KNN} 

(Temporal-Item-Frequency-Based User-KNN) is a neighbor-based approach developed for NBR. Unlike model-based methods, TIFU-KNN directly leverages the classical neighbor-based techniques to capture patterns in the data. The empirical findings from previous studies demonstrate that unseen patterns play a minor role in the target basket, justifying the adoption of neighbor-based methods.

TIFU-KNN introduces a novel approach to address the limitations of the Personalized Item Frequency (PIF) measure. While PIF provides valuable information, it fails to discriminate items with the same frequency, making it challenging to distinguish between them for neighbor searching and item selection in a kNN-based method. To overcome this limitation, the model incorporates the temporal dynamics of repeated purchases.

TIFU-KNN assigns group decay rate $r_g$ and within (group) decay rate $r_b$ to items based on their appearance at different time steps. Earlier appearances receive smaller weights, ensuring that recent purchases carry more significance in the final frequency calculation.

The methodology of TIFU-KNN consists of two key components: user similarity calculation and prediction. To calculate user similarity, the historical records of each user are aggregated into a single \textbf{user vector}. However, to account for drifting preferences over time, TIFU-KNN assigns higher importance to recent records than older ones. The model achieves this by employing hierarchical time-decayed weights.

For user similarity calculation, TIFU-KNN utilizes the Euclidean distance metric. The small distance between users indicates a higher similarity. The model searches for the k nearest neighbors for each target user.

In the prediction phase, TIFU-KNN combines the repeated and collaborative purchase components. The repeated purchase component is represented by the user's vector representation $\boldsymbol{u_t}$, while the collaborative purchase component is represented by the average vector of the user's nearest neighbors $\boldsymbol{u_n}$. The final prediction $\boldsymbol{P}$ is obtained through a linear combination of these two components, where a hyper-parameter $\alpha$ controls the balance between them:

\begin{equation} \label{eqn:5}
    \boldsymbol{P} = \alpha \cdot \boldsymbol{u_t} + (1 - \alpha) \cdot \boldsymbol{u_n}
\end{equation}

Based on the prediction, TIFU-KNN recommends the top $s$ items, corresponding to the largest entries in the prediction, as the next-basket recommendations for the target user.

\subsubsection{$\beta$-VAE} \label{sec:beta-vae}

The user vectors of TIFU-KNN are constructed heuristically and might have underlying latent patterns. Hence, we introduce another deep architecture for the NBR task making use of a Variational Autoencoder (VAE) network. Specifically, we implement the $\beta$-VAE variant \cite{HigginsMPBGBML17} with a simple Multi-layer Perceptron (MLP) form of encoder and decoder. The VAE network is trained to learn dense latent representations of dimension 128 for the user vectors which are sparse and of dimension equal to total items. We additionally train a second MLP decoder which we call \emph{Predictor} that predicts the next basket based on a target user's dense representation.

To evaluate the effect of collaborative patterns in NBR, we experiment with two variants of the model:
\begin{enumerate}
  \item We directly feed the target user's dense representation $\boldsymbol{u'_t}$ into the Predictor.
  \item We find out the $k$ nearest neighbors of the target user in the latent space and take an average over their dense vectors $\boldsymbol{u'_n}$. The two vectors are aggregated using Equation \ref{eqn:5} and then fed to the Predictor.
\end{enumerate}
We hypothesize that collaborative purchase patterns are important for NBR and the second form of implementation would outperform the first. The model and its hyperparameters are elaborated further in Appendix \ref{app:vae_imp}.

\subsection{Datasets}
In the original paper, the authors utilized four publicly available grocery shopping datasets: Dunnhumby\footnote{https://www.dunnhumby.com/careers/engineering/sourcefiles}, ValuedShopper\footnote{https://www.kaggle.com/competitions/acquire-valued-shoppers-challenge/data}, Instacart\footnote{https://www.kaggle.com/c/instacart-market-basket-analysis}, and TaFeng\footnote{https://www.kaggle.com/chiranjivdas09/ta-feng-grocery-dataset}. We used the train/test splits provided by the authors in their repository, where the last basket for each user was treated as test data, and all historical baskets were used as the training set. To ensure the presence of temporal patterns in the historical records, we followed the authors' approach and excluded customers with fewer than 3 baskets. Also, to ensure there is a collaborative pattern between users, we filtered out items that were purchased by less than 40 users for the Dunnhumby dataset and less than 5 for other datasets.

Additionally, to assess the generalizability of the findings on new datasets, we incorporated two more datasets into our study: Tmall\footnote{https://tianchi.aliyun.com/dataset/dataDetail?dataId=42} and Taobao\footnote{https://tianchi.aliyun.com/dataset/dataDetail?dataId=649}. These datasets are characterized by smaller basket sizes compared to the four previously mentioned datasets. To maintain comparability, we filtered out baskets smaller than four from the Tmall dataset. However, when we attempted the same filtering for the Taobao dataset, we were left with only a few hundred valid users. Consequently, we decided not to filter out smaller baskets in the Taobao dataset. Table \ref{tab:datasets_statistics} presents the statistics for all the datasets used in our study.

\begin{table}[!t]
  \centering
  \caption{Dataset statistics}
  \resizebox{\textwidth}{!}{
    \setlength{\tabcolsep}{3pt}
    \begin{tabular}{lrrrrrrr}
    \toprule
    \textbf{Dataset} & \textbf{\#Users} & {\textbf{\#Items}} & {\textbf{\#Baskets}} & {\textbf{Average}} & {\textbf{\#Baskets}} & {\textbf{Minimum}} & {\textbf{Maximum}} \\
     & & & & {\textbf{Basket Size}} & {\textbf{per user}} & {\textbf{Basket Size}} & {\textbf{Basket Size}} \\
    \midrule
    Instacart & 20k & 8k  & 307k & 8.98  & 15.47 & 1     & 89 \\
    Tmall & 20k & 24k & 83.5k & 4.05  & 4.07  & 1     & 65 \\
    Dunnhumby & 36k & 4.5k  & 289k & 7.36  & 7.98  & 1     & 2348 \\
    TaFeng & 14k & 12k & 93k & 6.28  & 6.71  & 1     & 109 \\
    Taobao & 4.6k & 32k & 16k & 2.88  & 3.48  & 2     & 49 \\
    ValuedShopper & 10k & 7.6k  & 598k & 8.72  & 59.87 & 1     & 128 \\
    \bottomrule
    \end{tabular}}
  \label{tab:datasets_statistics}
\end{table}

\subsection{Metrics}

In the original paper, two of the canonically used metrics in NBR literature are used to analyze the performance of the models, namely Recall@K and NDCG@K \cite{modeling-pif-hu}.

Other papers, such as \cite{li2023basket} consider additional metrics like the personalized hit ratio (PHR), and the mean of the Reciprocal Rank (MRR) \cite{shao2022systematical}. 

\begin{itemize}
    \item \textbf{MRR}: the Mean Reciprocal Rank measures how high the (first) relevant item is ranked. 
    \cite{shao2022systematical}
    \begin{equation}
        \mathrm{MRR} @ \mathrm{~K}=\frac{1}{N} \sum_{i=1}^N \frac{1}{q_i},
    \end{equation}
    where $q_i$ is defined as the ranking of the first item that appears in the ground truth in the top-K recommendation list, and $N$ is the number of test users \\
    
    \item \textbf{PHR}: the personalized-hit-ratio is defined as the ratio of users for whom at least one item in the predicted basket appears in the target basket. Defined as follows:
    \begin{equation}
        \mathrm{PHR} @ \mathrm{~K}=\frac{1}{N} \sum_{i=1}^N h r(i),
    \end{equation}
    where $hr(i)$ returns 1, if a user has at least 1 item in his target basket, appears in his prediction basket, and 0 otherwise\cite{shao2022systematical}.
\end{itemize}

In a similar fashion to \cite{li2023basket} we don't consider F1 and Precision metrics since we focus on the basket recommendation with a fixed basket size $K$, which means the Precision@$K$ and F1@$K$ are proportional to Recall@$K$ for each user. All metrics are considered at rank 10 and at rank 20 as these ranks are most commonly used in the current literature \cite{li2023basket} \cite{modeling-pif-hu} \cite{shao2022systematical}.

\subsection{Fairness} \label{sec:fairness}
In our study, we conducted a thorough examination of the fairness of the TIFU-KNN model. We performed an extensive analysis of fairness, considering three user characteristics: average basket size, the percentage of popular items in users' baskets, and the percentage of items never previously purchased by users in the test basket. As a main metric for comparison, we will report Recall@10 as it's widely used for NBR evaluation. \\

\textbf{Basket size}
We computed the average basket size for each user and examined whether the model's performance varied across different basket sizes. Our hypothesis postulates that the performance should be superior for larger baskets, as the model can leverage more information about the user's purchase patterns.

\textbf{Item popularity}
We identified the top 20\% of the most popular items purchased by all users. We then categorized each user based on the average number of these items present in their past baskets. Our hypothesis suggests that users who frequently purchase popular items should exhibit better model performance.

\textbf{Novelty}
For each user, we identified the set of items they had previously purchased and determined the percentage of items in the test basket that were not part of this set. Our hypothesis states that the model's performance should be lower for users with a higher proportion of unseen items in the test set.

\section{Experimental Setup/Implementation details } 
The authors of the original paper published their code \footnote{https://github.com/HaojiHu/TIFUKNN} however, the implementation is not very clear and modular therefore we continued our work based on the implementation from \cite{time-dependent-nbr-naumov}, as their code\footnote{https://github.com/sergunya17/time\_dependent\_nbr/tree/main} is better modularized, allowing for both an easier understanding of the code for any and a lower barrier to implementing changes. We further modified the codebase and adapted it to our needs by adding new datasets, new evaluation metrics, and changing the train test split methodology to be consistent with the original paper. Moreover, we added implementation for $\beta$-VAE architecture and removed unnecessary code. Our code is available here\footnote{https://github.com/pimpraat/RecSysProject}. In experiments, we used the hyperparameters reported in the original paper for Dunnhumby, Instacart, TaFeng, and ValuedShopper datasets. For Tmall and Taobao datasets we performed hyperparameter tuning over the parameter space to find the optimal values. The search was performed using the Optuna framework \cite{optuna_2019}, and the ranges used for parameters sampling can be found in Appendix \ref{app:hyperparameter_search}. In Table \ref{tab:optimal_hyperparameters} we report the optimal values for hyperparameters for all used datasets.

\setlength{\tabcolsep}{5pt}
\begin{table}[!t]
  \centering
  \caption{Optimal hyperparameters of TIFU-KNN for the datasets}
  \resizebox{\textwidth}{!}{
    \setlength{\tabcolsep}{5pt}
    \begin{tabular}{l r r r r r}
    \toprule
    \multicolumn{1}{p{6.335em}}{\textbf{Dataset}} & \multicolumn{1}{p{7em}}{\textbf{Num Nearest Neighbors}} & \multicolumn{1}{p{6.75em}}{\textbf{Within decay rate}} & \multicolumn{1}{p{6.5em}}{\textbf{Group decay rate}} & \multicolumn{1}{p{4em}}{\textbf{Group count}} & \multicolumn{1}{p{4em}}{\textbf{Alpha}} \\
    \midrule
    Instacart & 900   & 0.9   & 0.7   & 3     & 0.9 \\
    TaFeng & 300   & 0.9   & 0.7   & 7     & 0.7 \\
    Dunnhumby & 900   & 0.9   & 0.6   & 3     & 0.2 \\
    ValuedShopper & 300   & 1     & 0.6   & 7     & 0.7 \\
    Tmall & 100   & 0.6   & 0.8   & 18    & 0.7 \\
    Taobao & 300   & 0.6   & 0.8   & 10    & 0.1 \\
    \bottomrule
    \end{tabular}}
  \label{tab:optimal_hyperparameters}
\end{table}%

\section{Experimental Results} 
In this section we will report and elaborate on the results of the experiments, answering the research questions as defined in the previous section.

\subsection{RQ1: Reproduction of TIFU-KNN results} 
\begin{table}[!t]
\centering
\caption{Model performance on the original metrics, subscript indicating the absolute difference between reproduced results and reported results}
\label{table:_original_results}
\resizebox{\textwidth}{!}{
\setlength{\tabcolsep}{5pt}
\begin{tabular}{lccccc}
\toprule
Dataset & Model & R@10 & R@20 & NDCG@10 & NDCG@20 \\
\midrule
\multirow[c]{2}{*}{ValuedShopper} & TopPersonal & $ 0.2656_{+0.0547} $ & $ 0.3074_{+0.0105} $ & $ 0.2600_{+0.0472} $ & $ 0.2682_{+0.0138} $ \\
 & TIFUKNN & $ 0.2731_{+0.0569} $ & $ 0.3138_{+0.011} $ & $ 0.2663_{+0.0492} $ & $ 0.2740_{+0.0151} $ \\
\midrule
\multirow[c]{2}{*}{Tafeng} & TopPersonal & $ 0.1297_{+0.0593} $ & $ 0.1738_{+0.0535} $ & $ 0.1070_{0.0304} $ & $ 0.1219_{0.0323} $ \\
 & TIFUKNN & $0.1378_{+0.0077}$ & $0.1868_{+0.0058}$ & $0.1118_{+0.0107}$ & $0.1278_{+0.0107}$ \\
\midrule
\multirow[c]{2}{*}{Instacart} & TopPersonal & $0.3999_{+0.0573}$ & $0.4603_{-0.0049}$ & $0.4040_{+0.0422}$ & $0.4225_{+0.007}$ \\
 & TIFUKNN & $0.4308_{+0.0356}$ & $0.4962_{+0.0087}$ & $0.4247_{+0.0422}$ & $0.4478_{+0.0094}$ \\
\midrule
\multirow[c]{2}{*}{Dunnhumby} & TopPersonal & $0.2613_{-0.0556}$ & $0.3080_{+0.0714}$ & $0.2501_{+0.073}$ & $0.2616_{+0.06}$ \\
 & TIFUKNN & $0.2740_{+0.0653}$ & $0.3184_{+0.0492}$ & $0.2628_{+0.0645}$ & $0.2731_{+0.0429}$ \\
\midrule \midrule
\multirow[c]{2}{*}{Tmall} & TopPersonal & 0.1051 & 0.1262 & 0.0850 & 0.0939 \\
 & TIFUKNN & 0.1147 & 0.1412 & 0.0963 & 0.1071\\
\midrule
\multirow[c]{2}{*}{Taobao} & TopPersonal & 0.2671 & 0.2724 & 0.2279 & 0.2299 \\
 & TIFUKNN & 0.2701 & 0.2734 & 0.2498 & 0.2510 \\
\bottomrule
\end{tabular}}
\end{table}

Looking at the reproduction results, as can be seen in Table \ref{table:_original_results}, we observe for both models a slight (positive) deviation from the original results reported for both considered models. Since the parameters from the original paper itself were used, this difference in performance can be assumed to originate from differences in the pre-processing of the data and is supported by the observation that the difference increases at lower ranks.

However, even given these discrepancies/differences, it is just as in the original paper, observed that TIFU-KNN outperforms the TopPersonal model. We decided to compare the results only against the TopPersonal model because it is a widely used baseline that usually yields good performance.
\subsection{RQ2: Model performance on different datasets}

Results for additional datasets, namely Tmall and Taobao, along with the original datasets, are provided in Table \ref{table:_original_results}. The evaluation of various models' performance is based on the metrics R@10, R@20, NDCG@10, and NDCG@20.

Comparing these outcomes with those of other datasets reveals that the Tmall dataset exhibits similar performance to the TaFeng dataset. This similarity arises from their shared characteristics, as shown in Table \ref{tab:datasets_statistics}. Furthermore, although the Taobao dataset shares similar attributes with the Tmall and TaFeng datasets, its performance surpasses them and aligns with that of the ValuedShopper and Dunnhumby datasets. The latter datasets possess a substantially higher average number of baskets per user, which generally facilitates the model's ability to identify accurate purchase patterns.

Another noteworthy observation is the close proximity of metrics at 10 and 20 for the new datasets. This occurrence stems from these datasets having a smaller average basket size in comparison to the other datasets. Consequently, the metrics at 10 already encompass a significant portion of the items in the predicted baskets.

Overall, the Tmall and Taobao datasets introduce a distinctive set of challenges for NBR models, potentially necessitating further research and enhancements to attain improved performance.

\subsection{RQ3: Evaluation beyond recall and NDCG} 
\begin{table}[!t]
\centering
\caption{Model performance on the additional metrics}
\label{table:additional_results}
\setlength{\tabcolsep}{10pt}
\begin{tabular}{llrrrrr}
\toprule
Dataset & Model & MRR & PHR@10 & PHR@20 \\
\midrule
\multirow[c]{2}{*}{ValuedShopper} & TopPersonal & 0.4493 & 0.6834 & 0.7624 \\
 & TIFU-KNN & 0.4544 & 0.6923 & 0.7674 \\
\midrule
\multirow[c]{2}{*}{Tafeng} & TopPersonal & 0.1993 & 0.3819 & 0.4963 \\
 & TIFU-KNN & 0.2093 & 0.3998 & 0.5165 \\
\midrule
\multirow[c]{2}{*}{Instacart} & TopPersonal & 0.6335 & 0.8672 & 0.9169 \\
 & TIFU-KNN & 0.6346 & 0.8629 & 0.9083 \\
\midrule
\multirow[c]{2}{*}{Dunnhumby} & TopPersonal & 0.4162 & 0.6118 & 0.6883 \\
 & TIFU-KNN & 0.4355 & 0.6311 & 0.6993 \\
\midrule
\multirow[c]{2}{*}{Tmall} & TopPersonal & 0.1218 & 0.2319 & 0.2709 \\
 & TIFU-KNN & 0.1357 & 0.2423 & 0.2797 \\
\midrule
\multirow[c]{2}{*}{Taobao} & TopPersonal & 0.2282 & 0.3097 & 0.3162 \\
 & TIFU-KNN & 0.2561 & 0.3110 & 0.3149 \\
\bottomrule
\end{tabular}
\end{table}

Based on the experimental results for the additional metrics as defined earlier, several observations can be made:
\begin{itemize}
    \item Just as on the metrics used to analyze performance in the original paper, also on the Mean Reciprocal Rank metric the TIFU-KNN model outperforms the TopPersonal model, which solidates the findings that TIFU-KNN indeed outperforms the TopPersonal model (and likely the other models considered in the original paper too) on both the original and additional datasets and not just on the selected metrics.
    \item As expected, the personalized-hit-ratio for both considered models is similar, meaning the ratio of users for whom at least one item in the predicted basket is indeed in the target basket is almost identical. A possible extension for this metric, and a possibility for future work would be to define a hit for the PHR metric as having at least two items from the target basket in the predicted basket. These two values being similar for both models also indicates, again, the importance of PIF.
\end{itemize}

\subsection{RQ4: Fairness analysis of TIFU-KNN} Here we are presenting the results of our fairness study described in Section \ref{sec:fairness}.

\begin{figure}[h]
\centering
\includegraphics [width=\textwidth] {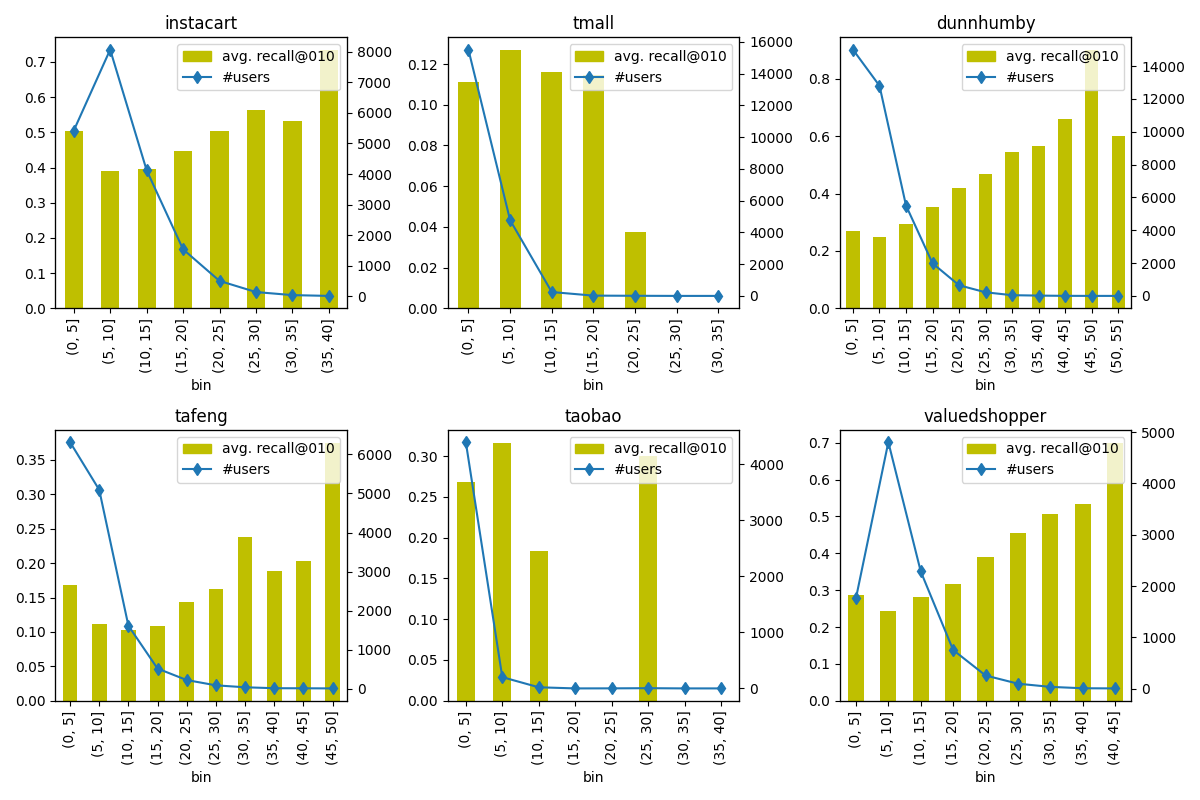}
\caption{Relationship between Model Performance and Basket Size. The yellow bars represent the values of Recall@10 on the left y-axis, while the blue line depicts the number of users on the right y-axis corresponding to different basket sizes along the x-axis.} 
\label{fig:fairness_basket_size}
\end{figure}

\textbf{Basket Size} Our hypothesis was partially validated as illustrated in Figure \ref{fig:fairness_basket_size}, wherein we observe improved performance for users with larger baskets in the ValuedShopper, TaFeng, Dunnhumby, and Instacart datasets. However, this trend is not observed in the case of the two new datasets, Tmall and Taobao. Furthermore, for the four original datasets, we observe a performance peak for basket sizes smaller than 5.

\begin{figure}[h]
\centering
\includegraphics [width=\textwidth] {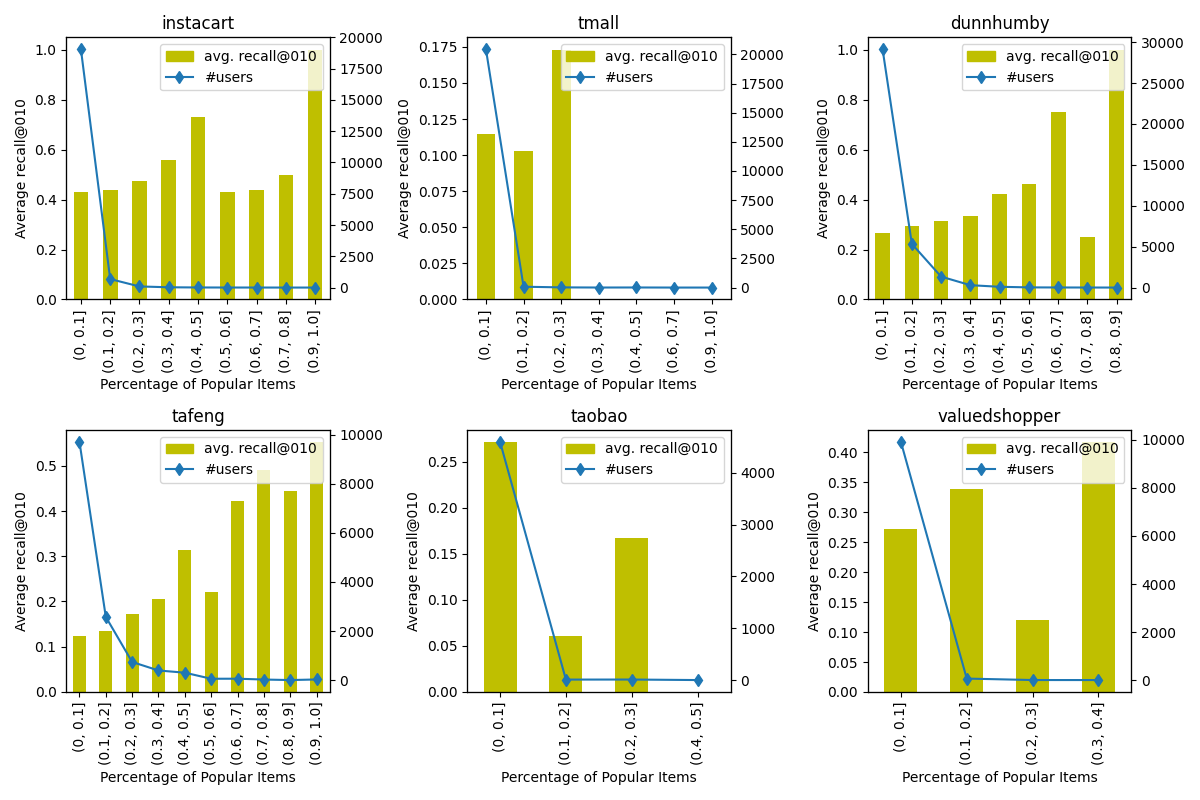}
\caption{Relationship between Model Performance and Item Popularity. The yellow bars represent the values of Recall@10 on the left y-axis, while the blue line depicts the number of users on the right y-axis corresponding to the percentage of popular items in users' baskets along the x-axis.} 
\label{fig:fairness_item_popularity}
\end{figure}

\textbf{Item popularity} Similarly to the basket size our hypothesis was partially confirmed. In Figure \ref{fig:fairness_item_popularity} we can see an evident trend for Dunnhumby, TaFeng, and Instacart datasets. However, this trend is not observed for other datasets. This indicates that bias toward item popularity depends on the characteristics of the datasets and that usually, the performance of the model is better when users are buying more popular products.

\begin{figure}[h]
\centering
\includegraphics [width=\textwidth] {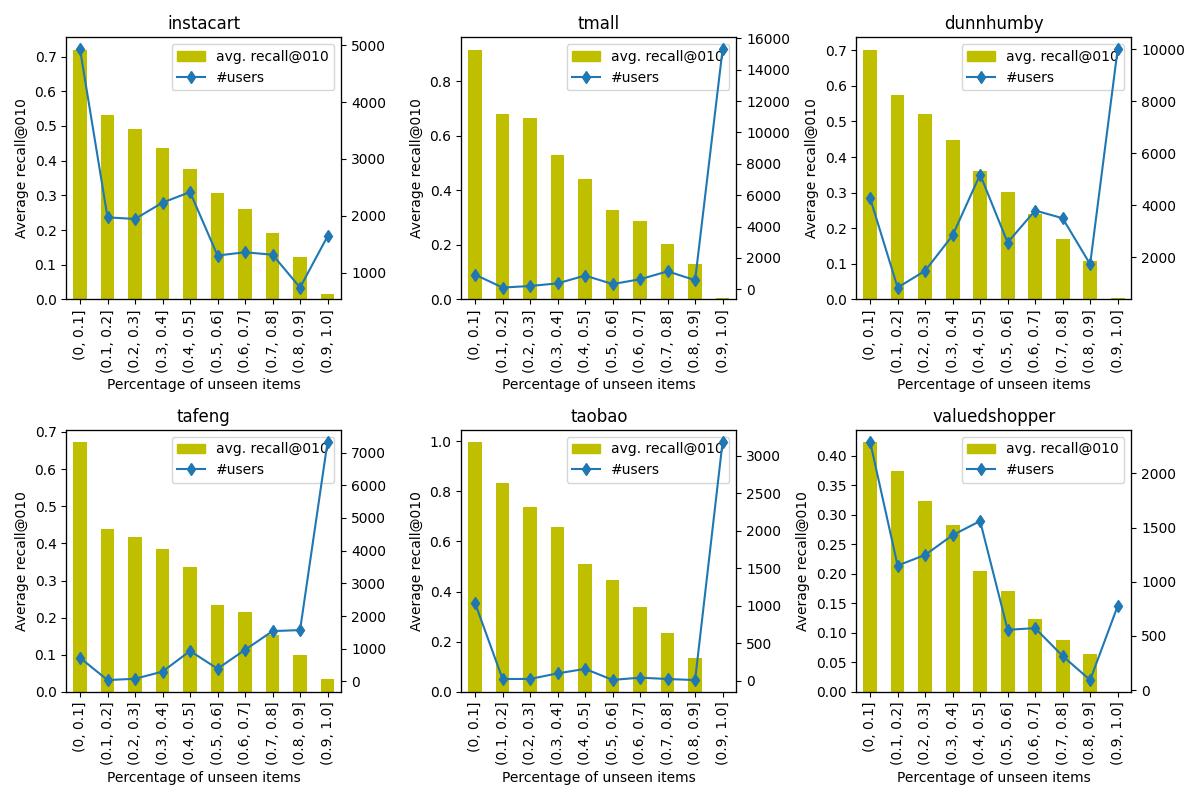}
\caption{Relationship between Model Performance and Novelty. The yellow bars represent the values of Recall@10 on the left y-axis, while the blue line depicts the number of users on the right y-axis corresponding to the percentage of unseen items in the test basket along the x-axis.} 
\label{fig:fairness_novelty}
\end{figure}

\textbf{Novelty} For novelty our hypothesis was confirmed for all datasets as depicted in Figure \ref{fig:fairness_novelty}a the trend is clearly visible. Moreover, we can see an interesting relation between performance and the percentage of unseen items when it is over 90\%. In that case, the value of recall is either 0 or close to 0 for all datasets which indicates that TIFU-KNN model cannot correctly predict the unseen items in the test basket. That's also the reason why TIFU-KNN has a rather poor performance on Tmall, Tafeng, and Taobao datasets as the proportion of unseen items in the test basket is high for the majority of users.

\subsection{RQ5: NBR with $\beta$-VAE}
The results for $\beta$-VAE architecture of the two types - with and without neighbor aggregation as described in Section \ref{sec:beta-vae} and Appendix \ref{app:vae_imp} are shown in Table \ref{table:_beta_vea_results}. The absolute values obtained for recall and NDCG are lower than the TopPersonal and TIFU-KNN methods for all datasets in consideration. This implementation may require finer architecture to handle sparse vectors, which we leave for future work. Even though this deep model performs poorly, the preliminary results are still intuitive. The relatively higher performance when k nearest neighbors are aggregated validates our hypothesis that collaborative patterns are important for NBR tasks.

Additionally, we examined the fairness of the $\beta$-VAE architecture and discovered that the deep model is more capable of predicting new items and performance is less dependent on the percentage of unseen items in the test basket. Results and a more detailed description are available in Appendix \ref{app:vae_novelty}.

\begin{table}[!t]
\centering
\caption{Comparison of $\beta$-VAE model performance when just the dense representation of users are used to predict the next basket (no kNN) and when dense representations of nearest k neighbors are aggregated (with kNN).}
\label{table:_beta_vea_results}
\resizebox{\textwidth}{!}{
\setlength{\tabcolsep}{8pt}
\begin{tabular}{llcccc}
\toprule
Dataset & Type & R@10 & R@20 & NDCG@10 & NDCG@20 \\
\midrule
\multirow[c]{2}{*}{ValuedShopper}
 & no kNN & 0.1253 & 0.1444 & 0.1243 & 0.1272 \\
 & with kNN (k=10) & 0.1258 & 0.1447 & 0.1249 & 0.1276 \\
\midrule
\multirow[c]{2}{*}{TaFeng}
 & no kNN & 0.1196 & 0.1433 & 0.1062 & 0.1148 \\
 & with kNN (k=10) & 0.1208 & 0.1445 & 0.1067 & 0.1154 \\
\midrule
\multirow[c]{2}{*}{Instacart} 
 & no kNN & 0.0531 & 0.0560 & 0.0529 & 0.0518 \\
 & with kNN (k=50) & 0.0964 & 0.1103 & 0.1016 & 0.1047 \\
\midrule
\multirow[c]{2}{*}{Dunnhumby}
 & no kNN & 0.0751 & 0.0733 & 0.0804 & 0.0751 \\
 & with kNN (k=20) & 0.0860 & 0.0921 & 0.0857 & 0.0838 \\
 \midrule
\multirow[c]{2}{*}{Tmall}
 & no kNN & 0.0004 & 0.0008 & 0.0002 & 0.0004 \\
 & with kNN (k=250) & 0.0201 & 0.0318 & 0.0155 & 0.0208 \\
\midrule
\multirow[c]{2}{*}{Taobao}
 & no kNN & 0.0034 & 0.0051 & 0.0021 & 0.0027 \\
 & with kNN (k=150) & 0.0031 & 0.0063 & 0.0018 & 0.0029 \\
\bottomrule
\end{tabular}}
\end{table}

\section{Discussion \& Conclusion} In this paper, we explored the application of the TIFU-KNN model for next-basket recommendation (NBR) and conducted an extensive analysis of its capabilities.

In our experiments, we reproduced the results reported in the original paper and evaluated the model's performance on different datasets and using other popular evaluation metrics. Although there were slight deviations in the reproduction results, the conclusions regarding the model's performance remained consistent. The Tmall and Taobao datasets presented unique challenges for NBR models, indicating the need for further research and enhancements.

Furthermore, we conducted a fairness analysis of the TIFU-KNN model. We examined three user characteristics, including average basket size, item popularity, and novelty. The results partially confirmed our hypotheses, indicating that performance varied based on these user characteristics. However, we observed some inconsistencies across different datasets, suggesting the influence of dataset-specific factors on fairness.

Lastly, we proposed the $\beta$-VAE architecture that makes use of the repeated purchase component in TIFU-KNN's user vectors to form dense latent representations. The recall and NDCG for the model are lower than TIFU-KNN and TopPersonal for which we reason the inability of the deep model to generate a logit vector with large dimensions. This is evident by the poor performances on Tmall and Taobao which have 24k and 32k items respectively, and hence form a very sparse vector. Using this model, we also show that collaborative purchase patterns between users and their neighbors are indeed important for NBR. Additionally, we also perform the novelty experiment for this model (see Appendix \ref{app:vae_novelty}) which shows that the deep model is more capable of predicting new items.

In conclusion, the TIFU-KNN model demonstrates competitive performance in next-basket recommendation tasks. It outperformed the dominant baseline model in terms of recall and NDCG metrics. However, fairness analysis revealed some variations in performance based on user characteristics, indicating the need for further investigation and potential model improvements to ensure equitable recommendations.

Future research could focus on refining the fairness evaluation framework, exploring additional user characteristics, and developing novel techniques to mitigate biases and improve fairness in NBR models. Additionally, improving $\beta$-VAE architecture and extensive analysis could provide further insights into its effectiveness and robustness.

\bibliographystyle{splncs04}
\bibliography{bibliography}

\newpage

\appendix

\section{Hyperparameter search ranges} \label{app:hyperparameter_search}
\begin{itemize}
    \item Num. Nearest Neighbors $k$: [100, 300, 500, 700, 900, 1100, 1300]
    \item Within decay rate $r_b$: [0.1, 0.2, 0.3, 0.4, 0.5, 0.6, 0.7, 0.8, 0.9, 1]
    \item Group decay rate $r_g$: [0.1, 0.2, 0.3, 0.4, 0.5, 0.6, 0.7, 0.8, 0.9, 1]
    \item The fusion weight $\alpha$: [0, 0.1, 0.2, 0.3, 0.4, 0.5, 0.6, 0.7, 0.8, 0.9, 1]
    \item The number of groups $m$: [2, 3, 4, 5, 6, 7, 8, 9, 10, 11, 12, 13, 14, 15, 16, 17, 18, 19, 20, 21, 22, 23]
\end{itemize}

\section{$\beta$-VAE Implementation} \label{app:vae_imp}

\begin{figure}[h]
\centering
\includegraphics [width=\textwidth] {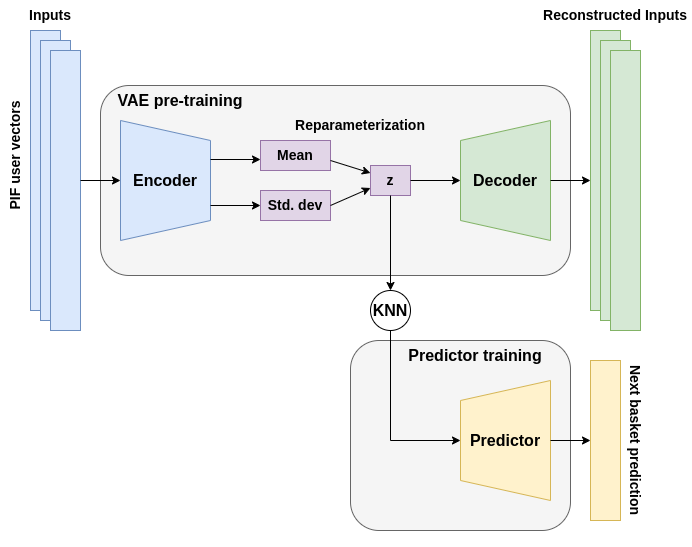}
\caption{Model architecture for the Next-basket Recommendation task using $\beta$-VAE and an MLP Predictor network.} 
\label{fig:pif_vae}
\end{figure}

Figure \ref{fig:pif_vae} shows the architecture for training a $\beta$-VAE model for the task of NBR. Hyperparameters for the model are further described in Table \ref{tab:vae_hyper_table}. The network is trained in two parts - First, the $\beta$-VAE is trained to learn latent dense representations of the user vectors. The objective is to minimize the loss function given by:

\begin{equation} \label{eqn:6}
    L(\theta, \phi, \beta) = - \E_{q_\phi(z|x)} \log{p_\theta(x|z)} + \beta D_{KL}(q_\phi(z|x)|| p_\theta(z)) 
\end{equation}

where, $q_\phi$ is the MLP encoder generating latent representation $z$ for an input $x$ and $p_\theta$ is the MLP decoder. $p_\theta(z)$ is the Gaussian prior and $D_{KL}$ is the KL-divergence. The Lagrangian multiplier $\beta$ acts as a disentanglement factor \cite{Burgess18} in the latent space. We use Mean Squared Error (MSE) as our reconstruction loss function (first term of Equation \ref{eqn:6}). We keep $\beta$ at 4 and train for 50 epochs with an initial learning rate of 0.005 and ReduceLRonPlateau strategy. The dense vectors of the users are then used to calculate their k nearest neighbors, where k is empirically found for each dataset. The final dense user vector is a $\alpha$-factored combination of the user vector from the encoder and the mean of its neighbor vectors similar to Equation \ref{eqn:5}.

Second, we train another Predictor network to output logits corresponding to item prediction for the next basket from which we select \emph{top-s} items. This network is trained for 50 epochs using MSE loss with q-hot true next-baskets for a subset of users. The initial learning rate was kept at 0.1 with the ReduceLRonPlateau strategy.

\begin{table}[!t]
    \small
    \centering
    \caption{Layer description and hyperparameters for components in the $\beta$-VAE model architecture. The last column denotes the activation function used. I/P Feats means Input Features. O/P Feats means output features.}
    \label{tab:vae_hyper_table}
    \resizebox{\textwidth}{!}{
    \begin{tabular}{l|cccc}
    \toprule
    \textbf{Component} & \textbf{Layer Type} & \textbf{Hyperparameters} & \textbf{Act. Func} \\
    \midrule 
    \multirow{4}{*}{Encoder} & \multirow{4}{*}{Linear} & I/P Feats: Total Items ; O/P Feats:1024 & LeakyReLU \\
    & &I/P Feats:1024 ; O/P Feats:512 & LeakyReLU \\
    & &I/P Feats:512 ; O/P Feats:256 & LeakyReLU \\
    & &I/P Feats:256 ; O/P Feats:128 x 2 & \\
    \midrule
    \multirow{3}{*}{\shortstack[l]{Reparameterisation\\ Trick}}&  & The output of encoder contains $\mu$ and log$\sigma$. \\
   & & The latent representation is then generated via & \\
      & &  $h = \mu(x)+ \sigma(x)\odot\epsilon, \epsilon \sim \mathcal{N}(0,1)$ \\
    \midrule
    \multirow{4}{*}{Decoder} & \multirow{4}{*}{Linear} & I/P Feats:128 ; O/P Feats:256 & LeakyReLU \\
    & &I/P Feats:256 ; O/P Feats:512 & LeakyReLU \\
    & &I/P Feats:512 ; O/P Feats:1024 & LeakyReLU \\
    & &I/P Feats:1024 ; O/P Feats: Total Items & Tanh \\
    \midrule \midrule
    \multirow{3}{*}{kNN} & & Algorithm: Brute Force & \\
    & & Similarity: Euclidean & \\
    & & Num. Neighbors: Dataset dependant & \\
    \midrule \midrule
    \multirow{4}{*}{Predictor} & \multirow{4}{*}{\shortstack[l]{Linear \\+ Dropout}} & I/P Feats:128 ; O/P Feats:256 & LeakyReLU \\
    & &I/P Feats:256 ; O/P Feats:512 & LeakyReLU \\
    & &I/P Feats:512 ; O/P Feats:1024 & LeakyReLU \\
    & &I/P Feats:1024 ; O/P Feats: Total Items & Tanh \\
    \bottomrule
    \end{tabular}}
\end{table}

\section{$\beta$-VAE Fairness in Novelty} \label{app:vae_novelty}

\begin{figure}[h]
\centering
\includegraphics [width=\textwidth] {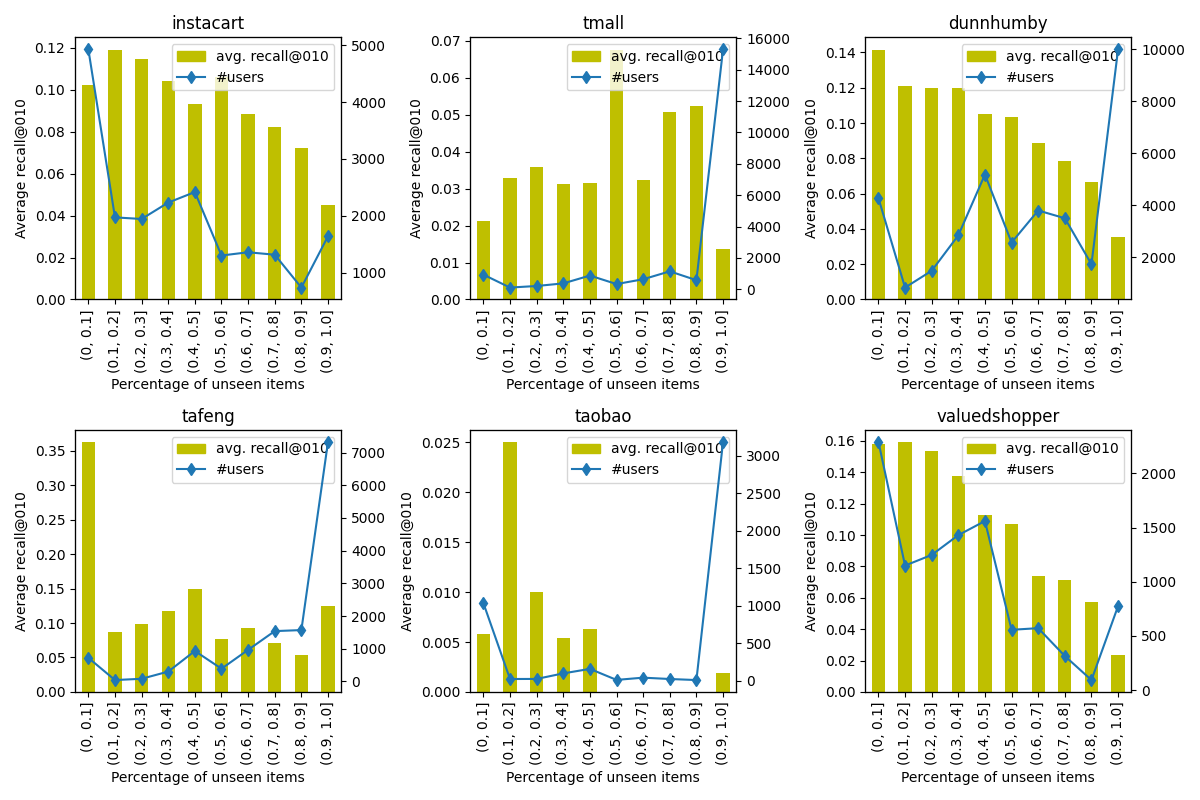}
\caption{Relationship between $\beta$-VAE Model Performance and Novelty. The yellow bars represent the values of Recall@10 on the left y-axis, while the blue line depicts the number of users on the right y-axis corresponding to the percentage of unseen items in the test basket along the x-axis.} 
\label{fig:vae_novelty_recall}
\end{figure}

Figure \ref{fig:vae_novelty_recall} shows the trend of the $\beta$-VAE model performance when users have new items in their next basket. Even though the plots for Dunnhumby and ValuedShopper show decreasing trends, the trends for TaFeng and Instacart are promising showing that the model is able to correctly predict more unseen items to the user. On comparing with trends obtained for the TIFU-KNN model in Figure \ref{fig:fairness_novelty}, we can observe that deep models are better at predicting new items than simple neighborhood-based models.

\end{document}